\documentclass[10pt,jcp,aip,twocolumn,showpacs,longbibliography,superscriptaddress]{revtex4-1}
\usepackage{epsfig,amsmath,amssymb,graphicx,color,calc,epstopdf}
\usepackage{color}
\usepackage{inputenc}

\newcommand{\br}{\mathbf{r}}
\newcommand{\bk}{\mathbf{k}}

\newcommand{\beq}{\begin{equation}}
\newcommand{\eeq}{\end{equation}}
\newcommand{\ba}{\begin{align}}
\newcommand{\ea}{\end{align}}

\makeatletter 
\@addtoreset{equation}{section}
\makeatother  

\usepackage{xr}

\begin{document}
\title{Microphase Equilibrium and Assembly Dynamics}
\author{Yuan Zhuang}
\affiliation{Department of Chemistry, Duke University, Durham,
	North Carolina 27708, USA}
\author{Patrick Charbonneau}
\affiliation{Department of Chemistry, Duke University, Durham,
	North Carolina 27708, USA}
\affiliation{Department of Physics, Duke University, Durham,
	North Carolina 27708, USA}

\begin{abstract}
Despite many attempts, ordered equilibrium microphases have yet to be obtained in experimental colloidal suspensions. The recent computation of the equilibrium phase diagram of a microscopic, particle--based microphase former (Zhuang et al., Phys.~Rev.~Lett. \textbf{116}, 098301 (2016)) has nonetheless found such mesoscale assemblies to be thermodynamically stable. Here, we consider their equilibrium and assembly dynamics. At intermediate densities above the order-disorder transition, we identify four different dynamical regimes and the structural changes that underlie the dynamical crossovers from one disordered regime to the next. Below the order-disorder transition, we also find that periodic lamellae are the most dynamically accessible of the periodic microphases. Our analysis thus offers a comprehensive view of the  disordered microphase dynamics and a route to the assembly of periodic microphases in a putative well-controlled, experimental system.
\end{abstract}


\maketitle
\paragraph*{Introduction --} Microphases replace liquid--gas coexistence in systems for which interparticle (short-range) attraction is frustrated by (long-range) repulsion, i.e., for SALR interactions~\cite{Brazovskii1975,Ciach2013,Zhuang2016c}. Beyond their mere aesthetic value, some of the resulting ordered and disordered structures of diblock copolymers have found uses as nanoscale templates and filters~\cite{Kim2010,Kataoka2001}. 
Colloidal analogues, however, have thus far only given rise to disordered microphases, such as clusters and gels~\cite{Campbell2005,Klix2010,Zhang2012}. Recent advances describing the equilibrium behavior of model colloidal microphase-formers~\cite{Zhuang2016,Zhuang2016a} confirm that such ordering should be thermodynamically possible~\cite{Ciach2013,Zhuang2016c}, but whether it is dynamically accessible remains an open question. 

Two main explanations have been proposed as to why periodic microphases might be difficult to access in model colloids. 
First, a glass-like dynamical slowdown in the amorphous microphase regime may hinder the assembly of periodic microphases. Different structural relaxation times grow as particles form clusters and eventually percolate at high enough particle density. In particular, it has been proposed that the structural relaxation time diverges near the percolation transition~\cite{Coniglio2004}. It has also been proposed that the system becomes glassy--once in the percolated regime--just above the order-disorder transition (ODT)~\cite{Geissler2004,Portmann2006,Schmalian2003,Haw2010}.
The super-Arrhenius slowdown of the relaxation dynamics would then severely hinder periodic microphase assembly~\cite{Geissler2004}.
Second, high free energy barriers between defective states may prevent periodic microphase from ordering, even modestly.  In diblock copolymers, microphase assembly sometimes gets stuck in locally stable structures that lack long-range order~\cite{Zhang2006,Tarzia2007,Campbell2005}. Upon (effective) cooling, a system would thus first form clusters or percolated clusters, and then rearrange into periodic microphases~\cite{Li2014,Tarzia2007,Tarzia2006,Uneyama2007}, with the latter severely hindered by the slow relaxation of topological defects~\cite{Groot1997,Riesch2014}. Because lamellae are at their most mobile just below the ODT, the ordering process should then be accelerated in that regime~\cite{Portmann2006}, and a slow annealing near the order-disorder transition be most likely to succeed~\cite{Li2014}. 

In either case, disentangling the microscopic contributions to periodic microphase assembly is essential to eventually exploiting their rich structural properties. In this Short Communication, we thus use a microphase-forming model for which the equilibrium phase diagram was recently determined~\cite{Zhuang2016,Zhuang2016a},
and analyze the equilibrium dynamics of disordered microphases and the assembly dynamics of lamellar microphases, in order to identify the origin of dynamical sluggishness in microphase assembly.

\paragraph*{Equilibrium Dynamics --} We simulate the local Monte Carlo (MC) dynamics of hard spherical particles of diameter $\sigma$ with a square-well linear (SWL) SALR pair interaction,
\begin{equation}\label{eq:swl}
u(r)=\begin{cases} 
\infty &\mbox{if } r\leq\sigma\\
-\varepsilon &\mbox{if } \sigma< r\leq\lambda\sigma \\
\xi\varepsilon(\kappa-r/\sigma) & \mbox{if } \lambda\sigma< r\leq\kappa\sigma \\
 0 & \mbox{if } r>\kappa\sigma 
 \end{cases},
\end{equation}
where $\epsilon$ is the attraction strength, $\xi$ controls the repulsion strength, $\lambda$ sets the attraction range,  and $\kappa$ sets the repulsion range. Note that from this point on, $\sigma$ implicitly sets the unit of length and $\epsilon$ that of energy. the For $\lambda=1.5$, $\xi=0.05$ and $\kappa=4$, the equilibrium phase diagram computed in Ref.~\onlinecite{Zhuang2016} identified a variety of equilibrium periodic microphases, including cluster crystals, cylinders and lamellae, at temperatures $T<T_{\mathrm{ODT}}$, while a structurally rich amorphous microphase regime was observed at $T>T_\mathrm{ODT}$. We consider here the equilibrium dynamics in the latter regime and the assembly dynamics in the former, by performing constant $NVT$  simulations with $N=1000$ particles at different $T$ and number density, $\rho=N/V$, within a cubic box of volume $V$, under periodic boundary conditions. 

\begin{figure}
\includegraphics[width=0.48\textwidth]{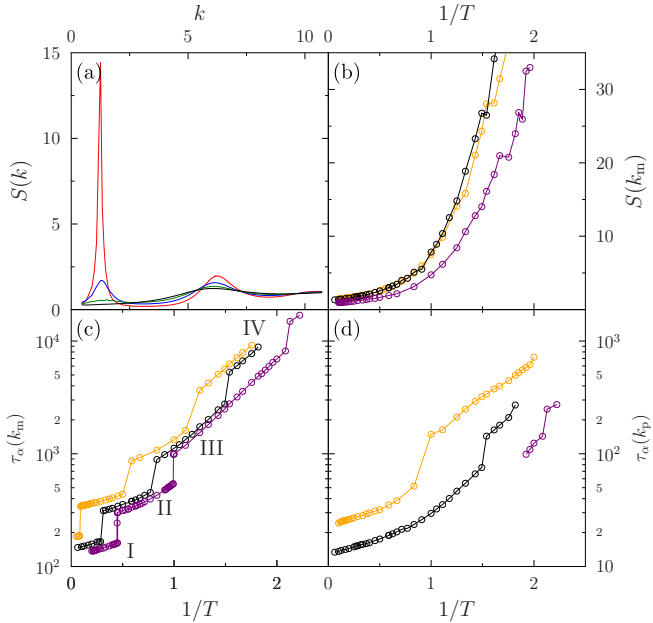}
    \caption{Structure factor in the disordered phase at $\rho=0.35$ for $T=0.55$~(red), $1.0$~(blue), $2.0$~(green) and $10.0$~(black), from top to bottom. The microphase signature at $k_\mathrm{m}$ appears as temperature is lowered.
	(b) Growth of the low-$k$ peak height, $S(k_\mathrm{m})$, with decreasing temperature for $\rho=0.3$ (purple), $0.35$ (black) and $0.4$ (orange), upon decreasing $T$, where $k_\mathrm{m}=1.19$, $1.25$ and $1.13$, respectively. The onset of the growth coincides with the crossover from region I to region II, but the subsequent growth is continuous within the numerical accuracy.
	(c) The structural relaxation time for microphase-scale features, $\tau_\alpha(k_\mathrm{m})$, at $\rho=0.3$ (purple), $0.35$ (black) and $0.4$ (orange), upon decreasing $T$ grows in marked steps. This feature delimits four different dynamical regime.
	(d) The structural relaxation time for particle-scale features, $\tau_\alpha(k_\mathrm{p})$, also grows, but its only marked increase occurs around the crossover from region III to region IV. Note that for $\rho=0.3$ at high temperatures, the dynamics is too fast for our simulations to precisely this quantity.}
\label{fig:dynamic_slow}
\end{figure}

The structure factor, $
S(k) = N^{-1}\sum_{ij}^N \left\langle\mathrm{e}^{i \mathbf{k}\cdot(\mathbf{r}_i-\mathbf{r}_j)} \right\rangle$, with wavevector $\mathbf{k}=\frac{2\pi}{\sqrt[1/3]{V}}\mathbf{n}$ for $\mathbf{n}\in\mathbb{Z}^3$, synthesizes the spatial organization of the system. Its peak at $k_\mathrm{p}\approx 2\pi$ captures dominant particle-scale features, while its low-$k$ peak at $k_\mathrm{m}\approx 1.2$ characterizes microphase-scale features, when they exist~(Fig.~\ref{fig:dynamic_slow}~(a) and (b)).  The structural relaxation time, $\tau_\alpha(k)$, extracted from the characteristic decay time of the self-intermediate scattering function
\begin{equation}\label{eq:corr}
F_\mathrm{s}(t;k) = \frac{1}{N}\sum_{i}\left\langle\exp\left[i\bk\cdot\left(\br_i(t)-\br_i(0)\right)\right]\right\rangle,
\end{equation}
probes the dynamical relaxation of one or the other type of features, depending on the chosen $k$. In practice, wavevectors within 5\% of $k_\mathrm{m}$ and within 2\% of $k_\mathrm{p}$ are used, so as to reduce statistical noise.

\begin{figure}
	\includegraphics[width=0.42\textwidth]{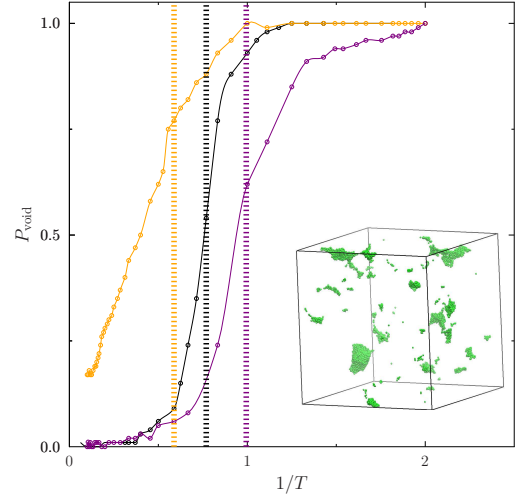}
	\caption{
Void percolation probability (concurrently in all three spatial dimensions), $P_\mathrm{void}$, obtained from inserting particles of diameter $0.01\sigma$ in a system of $N=1000$ particle at $\rho=0.3$~(purple), $0.35$~(black) and $0.4$~(orange). The crossover regions between regime II and III obtained in Fig.~\ref{fig:dynamic_slow}~(c) (dashed lines) roughly correspond to the onset of void percolation. (inset) Sample convex hulls obtained from inserted particles of diameter $0.05$ into an equilibrium configuration at $T=2.0$ and $\rho=0.4$.}
\label{fig:void}
\end{figure}

Rather than a glass-like super-Arrhenius-scaling upon cooling, $\tau_\alpha(k_\mathrm{m})$ hints at the existence of four distinct Arrhenius-like scaling regimes~(Fig.~\ref{fig:dynamic_slow}~(c)). We label them I-IV, from high to low $T$. Although no phase transition separates these different regimes, the relative sharpness of the crossovers that separate them suggests that marked structural changes nonetheless accompany the dynamical slowdown. The precise nature of these crossovers is, however, not immediately obvious. 

In regime I,  structural relaxation is fast and the low-$k$ behavior of $S(k)$ is featureless, which is consistent with the system being simple-fluid--like. The crossover to regime II, by contrast, also roughly corresponds to the onset of the growth of $S(k_\mathrm{m})$ (Fig.~\ref{fig:dynamic_slow}~(b)), while the single-particle dynamics remains fast. This suggests that amorphous, fluid microphase features then emerge. The crossover from regime III to IV is also rather straightforward to characterize. A jump in both $\tau_\alpha(k_\mathrm{m})$ and $\tau_\alpha(k_\mathrm{p})$ takes place in that temperature range, while the other crossovers only affected $\tau_\alpha(k_\mathrm{m})$ (Fig.~\ref{fig:dynamic_slow}(d)). This particular slowdown is thus akin to individual particles condensing. The deceleration of the microphase-scale relaxation then directly follows from the slowdown of the particle-scale dynamics.

Because it is not accompanied by significant changes to the structure factor nor by particle clustering, the crossover from regimes II to III is more confounding. Marked structural changes, if they do take place, must thus somehow paradoxically be rather subtle. As a potential structural explanation, we consider possible changes to the void structure, which would affect $S(k)$ only indirectly and may otherwise go unnoticed. To do so, we  first randomly insert $10^5$ small hard spheres within a set of $100$ structurally independent equilibrated configurations, i.e., separated by at least $200\tau_\mathrm{\alpha}(k_\mathrm{m})$. We then extracted the various three-dimensional convex hulls formed by the union of the inserted spheres~(Figure~\ref{fig:void} inset), and determined the onset of void percolation. Because the void properties extracted this way depend on the diameter of the inserted spheres, we chose spheres smaller than the cavity size contained within four touching spheres, but not much smaller than the hole between three coplanar spheres. In practice, using particles over the range of diameters $0.01\sigma$ to $0.05\sigma$ reveals no substantial difference, which  validates the robustness of the findings. Remarkably, the onset of void percolation does roughly fall in line with the crossover from II to III (Fig~\ref{fig:void}). The associated dynamical slowdown thus reflects the morphological changes in the void structure as it goes from compact to percolating.

\begin{figure}
    \includegraphics[width=0.4\textwidth]{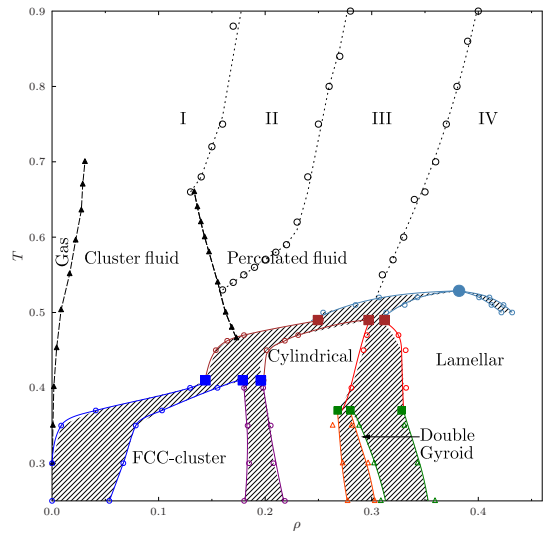}
    \caption{The different dynamical regimes in the disordered phase are superimposed on the equilibrium ordered microphase results of the SWL model from Ref.~\cite{Zhuang2016} (solid lines and shaded coexistence regions). The dynamical crossover lines (dotted lines) are obtained from a dynamical analysis as in Fig.~\ref{fig:dynamic_slow}, while the clustering and cluster percolation lines (dashed) are from Ref.~\cite{Zhuang2016}.} \label{fig:phase_diagram}
\end{figure}

Superimposing the three crossovers with the equilibrium phase information in Fig.~\ref{fig:phase_diagram} provides an overall view of the microphase morphology and dynamics.
In the high-temperature regime~(I), the system is akin to a homogeneous, hard-sphere-like simple fluid. The SALR contribution to the interaction plays essentially no role. 
Regime~(II) sees the emergence of void clustering, which in a sense complements the formation of particle clusters as density increases. This clustering is, however, seemingly distinct from that observed in SALR systems at even higher densities~\cite{Zhuang2016c,Lindquist2016,Lindquist2017}
Upon further lowering temperature, the void clusters become wormlike and eventually percolate, thus giving rise to an amorphous, bicontinuous microphase structure (III). The formation of these structures is again complementary to the particle clusters becoming wormlike and eventually percolating upon increasing the particle density. Throughout regimes I-III, the local particle dynamics remains fluid-like, but regime IV shows a marked local slowdown, akin to a gas-solid condensation. This last crossover indeed occurs around $T=0.6$ and takes place at higher temperatures as $\rho$ increases. 
Because the condensed system nonetheless preserves a lot of surface area the dynamics cannot actually freeze. Rearrangements carry thus through the bond breaking of surface particles. 

\begin{figure}
	\includegraphics[width=0.4\textwidth]{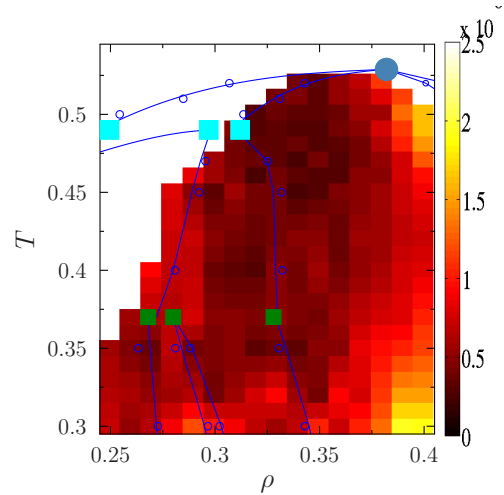}
	\caption{Heat map of the kinetic ease of assembly of periodic lamellae, with the simulation time given in multiples of $100$ MC steps. Superimposing the result with the equilibrium phase diagram indicates that assembly is most dynamically facile near the ODT, where defect annealing is fastest and the first-order transition is weakest. At slightly higher densities, assembly slows downs markedly. No cylinder nor gyroid phases assembled, although a mix of cylinders and lamellae do form outside of the equilibrium regime for pure lamellae.}\label{fig:time}
\end{figure}

\paragraph*{Assembly Dynamics --} 
Having established that no glass-like slowing down is observed in the amorphous microphase regime, we consider whether periodic microphase assembly is dynamically accessible and to what extent. To evaluate the propensity of lamellae to assemble, $40$ independent MC simulations are run for $2\times10^6$ MC steps after quenching equilibrated configurations at $T=0.55>T_\mathrm{ODT}$ to various state points within the periodic microphase regime.
Although the amorphous and the periodic microphases share a same $k_\mathrm{m}$, the peak intensity, $S(k_\mathrm{m})$ markedly differs between the two. The former scales as $N^0$ and the latter as $N^1$. We thus use the time at which $S(k_\mathrm{m})>30$ to denote the formation of periodic lamellae. Note, however that this criterion does distinguish lamellae from cylindrical or cubic crystal clusters which all display a large low-$k$ peak. A visual inspection of the configurations was thus also done in the regime where these other structures  could appear. 

From this analysis we obtain a lower bound on the dynamical ease with which lamellae self assemble. Superimposing this map onto the phase diagram clearly shows that the lamellar phase forms more robustly and efficiently just below the ODT (see Fig.~\ref{fig:time}). The difference between the regimes where assembly is facile and where it is hard reveals that a slowdown by a factor of at least $10$. For instance, at densities slightly higher than the ODT density assembly is much slower, while at densities slightly lower the assembly time is nearly constant. Further lowering density below the lamellar regime reveals the rise of connected lamellae, which mix cylinders and lamellae. Remarkably, over the simulation timescale explored, no hint of cylindrical or double-gyroid phase were observed even though they should be thermodynamically stable at some of the state points we studied. High free energy barriers for
annealing defects in those morphologies might then indeed be the culprit. 

\paragraph*{Conclusion --} We have identified four dynamical regimes within the disordered phase of a model SALR system. Within each of these regimes the growth of the relaxation time is Arrhenius-like, $\tau_\mathrm{\alpha}\sim\exp(\beta\Delta E_\mathrm{a})$, but the associated relaxation energy $\Delta E_\mathrm{a}$ increases from one regime to another. From regime I to IV the quantity increases by a factor of about $50$, which suggests that structural relaxations involve increasingly large regions as temperature decreases. Taking the overall disordered regime at once, however, sees the ``super-Arrhenius'' dynamics slowing down by barely more than two decades. This dynamical fragility of sort is therefore distinct from a typical glassy slowdown.

Our study of the microphase assembly dynamics of the SWL system further demonstrates that nothing fundamentally prevents periodic lamellae from forming in colloidal-like systems. Despite the complex dynamics of the disordered regime, the self-assembly of lamellae remains rather fast. The absence of periodic microphases in colloidal experiments~\cite{Campbell2005,Klix2010} is therefore likely not limited by the nature of the order-disorder transition, but rather from the specifics of previously studied systems, such as their short-range attraction, the low densities considered, or the limited control thus far achieved over experimental conditions. The challenge of obtaining a colloidal realization should thus be surmountable.

\begin{acknowledgments}
We acknowledge support from the National Science Foundation Grant no. NSF DMR-1055586 and from the Materials Research Science and Engineering Centers~(DMR-1121107). This work used the Extreme Science and Engineering Discovery Environment (XSEDE), which is supported by National Science Foundation grant number ACI-1548562~\cite{6866038}.
\end{acknowledgments}

   \bibliography{./references}

\begin{thebibliography}{25}%
\makeatletter
\providecommand \@ifxundefined [1]{%
 \@ifx{#1\undefined}
}%
\providecommand \@ifnum [1]{%
 \ifnum #1\expandafter \@firstoftwo
 \else \expandafter \@secondoftwo
 \fi
}%
\providecommand \@ifx [1]{%
 \ifx #1\expandafter \@firstoftwo
 \else \expandafter \@secondoftwo
 \fi
}%
\providecommand \natexlab [1]{#1}%
\providecommand \enquote  [1]{``#1''}%
\providecommand \bibnamefont  [1]{#1}%
\providecommand \bibfnamefont [1]{#1}%
\providecommand \citenamefont [1]{#1}%
\providecommand \href@noop [0]{\@secondoftwo}%
\providecommand \href [0]{\begingroup \@sanitize@url \@href}%
\providecommand \@href[1]{\@@startlink{#1}\@@href}%
\providecommand \@@href[1]{\endgroup#1\@@endlink}%
\providecommand \@sanitize@url [0]{\catcode `\\12\catcode `\$12\catcode
  `\&12\catcode `\#12\catcode `\^12\catcode `\_12\catcode `\%12\relax}%
\providecommand \@@startlink[1]{}%
\providecommand \@@endlink[0]{}%
\providecommand \url  [0]{\begingroup\@sanitize@url \@url }%
\providecommand \@url [1]{\endgroup\@href {#1}{\urlprefix }}%
\providecommand \urlprefix  [0]{URL }%
\providecommand \Eprint [0]{\href }%
\providecommand \doibase [0]{http://dx.doi.org/}%
\providecommand \selectlanguage [0]{\@gobble}%
\providecommand \bibinfo  [0]{\@secondoftwo}%
\providecommand \bibfield  [0]{\@secondoftwo}%
\providecommand \translation [1]{[#1]}%
\providecommand \BibitemOpen [0]{}%
\providecommand \bibitemStop [0]{}%
\providecommand \bibitemNoStop [0]{.\EOS\space}%
\providecommand \EOS [0]{\spacefactor3000\relax}%
\providecommand \BibitemShut  [1]{\csname bibitem#1\endcsname}%
\let\auto@bib@innerbib\@empty
\bibitem [{\citenamefont {Brazovskii}(1975)}]{Brazovskii1975}%
  \BibitemOpen
  \bibfield  {author} {\bibinfo {author} {\bibfnamefont {S.~A.}\ \bibnamefont
  {Brazovskii}},\ }\bibfield  {title} {\enquote {\bibinfo {title} {Phase
  transition of an isotropic system to a nonuniform state},}\ }\href@noop {}
  {\bibfield  {journal} {\bibinfo  {journal} {Sov. Phys. JETP}\ }\textbf
  {\bibinfo {volume} {41}},\ \bibinfo {pages} {85--89} (\bibinfo {year}
  {1975})}\BibitemShut {NoStop}%
\bibitem [{\citenamefont {Ciach}, \citenamefont {Pekalski},\ and\ \citenamefont
  {G{\'o}{\'z}d{\'z}}(2013)}]{Ciach2013}%
  \BibitemOpen
  \bibfield  {author} {\bibinfo {author} {\bibfnamefont {A.}~\bibnamefont
  {Ciach}}, \bibinfo {author} {\bibfnamefont {J.}~\bibnamefont {Pekalski}}, \
  and\ \bibinfo {author} {\bibfnamefont {W.}~\bibnamefont
  {G{\'o}{\'z}d{\'z}}},\ }\bibfield  {title} {\enquote {\bibinfo {title}
  {Origin of similarity of phase diagrams in amphiphilic and colloidal systems
  with competing interactions},}\ }\href@noop {} {\bibfield  {journal}
  {\bibinfo  {journal} {Soft Matter}\ }\textbf {\bibinfo {volume} {9}},\
  \bibinfo {pages} {6301--6308} (\bibinfo {year} {2013})}\BibitemShut {NoStop}%
\bibitem [{\citenamefont {Zhuang}\ and\ \citenamefont
  {Charbonneau}(2016{\natexlab{a}})}]{Zhuang2016c}%
  \BibitemOpen
  \bibfield  {author} {\bibinfo {author} {\bibfnamefont {Y.}~\bibnamefont
  {Zhuang}}\ and\ \bibinfo {author} {\bibfnamefont {P.}~\bibnamefont
  {Charbonneau}},\ }\bibfield  {title} {\enquote {\bibinfo {title} {Recent
  advances in the theory and simulation of model colloidal microphase
  formers},}\ }\href@noop {} {\bibfield  {journal} {\bibinfo  {journal} {J.
  Chem. Phys. B}\ }\textbf {\bibinfo {volume} {120}},\ \bibinfo {pages}
  {7775--7782} (\bibinfo {year} {2016}{\natexlab{a}})}\BibitemShut {NoStop}%
\bibitem [{\citenamefont {Kim}, \citenamefont {Park},\ and\ \citenamefont
  {Hinsberg}(2010)}]{Kim2010}%
  \BibitemOpen
  \bibfield  {author} {\bibinfo {author} {\bibfnamefont {H.~C.}\ \bibnamefont
  {Kim}}, \bibinfo {author} {\bibfnamefont {S.~M.}\ \bibnamefont {Park}}, \
  and\ \bibinfo {author} {\bibfnamefont {W.~D.}\ \bibnamefont {Hinsberg}},\
  }\bibfield  {title} {\enquote {\bibinfo {title} {Block copolymer based
  nanostructures: Materials, processes, and applications to electronics},}\
  }\href@noop {} {\bibfield  {journal} {\bibinfo  {journal} {Chem. Rev.}\
  }\textbf {\bibinfo {volume} {110}},\ \bibinfo {pages} {146--177} (\bibinfo
  {year} {2010})}\BibitemShut {NoStop}%
\bibitem [{\citenamefont {Kataoka}, \citenamefont {Harada},\ and\ \citenamefont
  {Nagasaki}(2001)}]{Kataoka2001}%
  \BibitemOpen
  \bibfield  {author} {\bibinfo {author} {\bibfnamefont {K.}~\bibnamefont
  {Kataoka}}, \bibinfo {author} {\bibfnamefont {A.}~\bibnamefont {Harada}}, \
  and\ \bibinfo {author} {\bibfnamefont {Y.}~\bibnamefont {Nagasaki}},\
  }\bibfield  {title} {\enquote {\bibinfo {title} {Block copolymer micelles for
  drug delivery: Design, characterization and biological significance},}\
  }\href@noop {} {\bibfield  {journal} {\bibinfo  {journal} {Adv. Drug. Deliv.
  Rev.}\ }\textbf {\bibinfo {volume} {47}},\ \bibinfo {pages} {113--131}
  (\bibinfo {year} {2001})}\BibitemShut {NoStop}%
\bibitem [{\citenamefont {Campbell}\ \emph {et~al.}(2005)\citenamefont
  {Campbell}, \citenamefont {Anderson}, \citenamefont {van Duijneveldt},\ and\
  \citenamefont {Bartlett}}]{Campbell2005}%
  \BibitemOpen
  \bibfield  {author} {\bibinfo {author} {\bibfnamefont {A.~I.}\ \bibnamefont
  {Campbell}}, \bibinfo {author} {\bibfnamefont {V.~J.}\ \bibnamefont
  {Anderson}}, \bibinfo {author} {\bibfnamefont {J.~S.}\ \bibnamefont {van
  Duijneveldt}}, \ and\ \bibinfo {author} {\bibfnamefont {P.}~\bibnamefont
  {Bartlett}},\ }\bibfield  {title} {\enquote {\bibinfo {title} {Dynamical
  arrest in attractive colloids: The effect of long-range repulsion},}\
  }\href@noop {} {\bibfield  {journal} {\bibinfo  {journal} {Phys. Rev. Lett.}\
  }\textbf {\bibinfo {volume} {94}},\ \bibinfo {pages} {208301} (\bibinfo
  {year} {2005})}\BibitemShut {NoStop}%
\bibitem [{\citenamefont {Klix}, \citenamefont {Royall},\ and\ \citenamefont
  {Tanaka}(2010)}]{Klix2010}%
  \BibitemOpen
  \bibfield  {author} {\bibinfo {author} {\bibfnamefont {C.~L.}\ \bibnamefont
  {Klix}}, \bibinfo {author} {\bibfnamefont {C.~P.}\ \bibnamefont {Royall}}, \
  and\ \bibinfo {author} {\bibfnamefont {H.}~\bibnamefont {Tanaka}},\
  }\bibfield  {title} {\enquote {\bibinfo {title} {Structural and dynamical
  features of multiple metastable glassy states in a colloidal system with
  competing interactions},}\ }\href@noop {} {\bibfield  {journal} {\bibinfo
  {journal} {Phys. Rev. Lett.}\ }\textbf {\bibinfo {volume} {104}},\ \bibinfo
  {pages} {165702} (\bibinfo {year} {2010})}\BibitemShut {NoStop}%
\bibitem [{\citenamefont {Zhang}\ \emph {et~al.}(2012)\citenamefont {Zhang},
  \citenamefont {Klok}, \citenamefont {Hans~Tromp}, \citenamefont
  {Groenewold},\ and\ \citenamefont {Kegel}}]{Zhang2012}%
  \BibitemOpen
  \bibfield  {author} {\bibinfo {author} {\bibfnamefont {T.~H.}\ \bibnamefont
  {Zhang}}, \bibinfo {author} {\bibfnamefont {J.}~\bibnamefont {Klok}},
  \bibinfo {author} {\bibfnamefont {R.}~\bibnamefont {Hans~Tromp}}, \bibinfo
  {author} {\bibfnamefont {J.}~\bibnamefont {Groenewold}}, \ and\ \bibinfo
  {author} {\bibfnamefont {W.~K.}\ \bibnamefont {Kegel}},\ }\bibfield  {title}
  {\enquote {\bibinfo {title} {Non-equilibrium cluster states in colloids with
  competing interactions},}\ }\href@noop {} {\bibfield  {journal} {\bibinfo
  {journal} {Soft Matter}\ }\textbf {\bibinfo {volume} {8}},\ \bibinfo {pages}
  {667--672} (\bibinfo {year} {2012})}\BibitemShut {NoStop}%
\bibitem [{\citenamefont {Zhuang}, \citenamefont {Zhang},\ and\ \citenamefont
  {Charbonneau}(2016)}]{Zhuang2016}%
  \BibitemOpen
  \bibfield  {author} {\bibinfo {author} {\bibfnamefont {Y.}~\bibnamefont
  {Zhuang}}, \bibinfo {author} {\bibfnamefont {K.}~\bibnamefont {Zhang}}, \
  and\ \bibinfo {author} {\bibfnamefont {P.}~\bibnamefont {Charbonneau}},\
  }\bibfield  {title} {\enquote {\bibinfo {title} {Equilibrium phase behavior
  of a continuous-space microphase former},}\ }\href@noop {} {\bibfield
  {journal} {\bibinfo  {journal} {Phys. Rev. Lett.}\ }\textbf {\bibinfo
  {volume} {116}},\ \bibinfo {pages} {098301} (\bibinfo {year}
  {2016})}\BibitemShut {NoStop}%
\bibitem [{\citenamefont {Zhuang}\ and\ \citenamefont
  {Charbonneau}(2016{\natexlab{b}})}]{Zhuang2016a}%
  \BibitemOpen
  \bibfield  {author} {\bibinfo {author} {\bibfnamefont {Y.}~\bibnamefont
  {Zhuang}}\ and\ \bibinfo {author} {\bibfnamefont {P.}~\bibnamefont
  {Charbonneau}},\ }\bibfield  {title} {\enquote {\bibinfo {title} {Equilibrium
  phase behavior of the square-well linear microphase-forming model},}\
  }\href@noop {} {\bibfield  {journal} {\bibinfo  {journal} {J. Phys. Chem.
  B,}\ }\textbf {\bibinfo {volume} {120}},\ \bibinfo {pages} {6178--6188}
  (\bibinfo {year} {2016}{\natexlab{b}})}\BibitemShut {NoStop}%
\bibitem [{\citenamefont {Coniglio}\ \emph {et~al.}(2004)\citenamefont
  {Coniglio}, \citenamefont {Arcangelis}, \citenamefont {Gado}, \citenamefont
  {Fierro},\ and\ \citenamefont {Sator}}]{Coniglio2004}%
  \BibitemOpen
  \bibfield  {author} {\bibinfo {author} {\bibfnamefont {A.}~\bibnamefont
  {Coniglio}}, \bibinfo {author} {\bibfnamefont {L.~D.}\ \bibnamefont
  {Arcangelis}}, \bibinfo {author} {\bibfnamefont {E.~D.}\ \bibnamefont
  {Gado}}, \bibinfo {author} {\bibfnamefont {A.}~\bibnamefont {Fierro}}, \ and\
  \bibinfo {author} {\bibfnamefont {N.}~\bibnamefont {Sator}},\ }\bibfield
  {title} {\enquote {\bibinfo {title} {Percolation, gelation and dynamical
  behaviour in colloids},}\ }\href
  {http://stacks.iop.org/0953-8984/16/i=42/a=002} {\bibfield  {journal}
  {\bibinfo  {journal} {J. Phys.: Condens. Matter}\ }\textbf {\bibinfo {volume}
  {16}},\ \bibinfo {pages} {S4831} (\bibinfo {year} {2004})}\BibitemShut
  {NoStop}%
\bibitem [{\citenamefont {Geissler}\ and\ \citenamefont
  {Reichman}(2004)}]{Geissler2004}%
  \BibitemOpen
  \bibfield  {author} {\bibinfo {author} {\bibfnamefont {P.~L.}\ \bibnamefont
  {Geissler}}\ and\ \bibinfo {author} {\bibfnamefont {D.~R.}\ \bibnamefont
  {Reichman}},\ }\bibfield  {title} {\enquote {\bibinfo {title} {Nature of slow
  dynamics in a minimal model of frustration-limited domains},}\ }\href@noop {}
  {\bibfield  {journal} {\bibinfo  {journal} {Phys. Rev. E}\ }\textbf {\bibinfo
  {volume} {69}},\ \bibinfo {pages} {021501} (\bibinfo {year}
  {2004})}\BibitemShut {NoStop}%
\bibitem [{\citenamefont {Portmann}, \citenamefont {Vaterlaus},\ and\
  \citenamefont {Pescia}(2006)}]{Portmann2006}%
  \BibitemOpen
  \bibfield  {author} {\bibinfo {author} {\bibfnamefont {O.}~\bibnamefont
  {Portmann}}, \bibinfo {author} {\bibfnamefont {A.}~\bibnamefont {Vaterlaus}},
  \ and\ \bibinfo {author} {\bibfnamefont {D.}~\bibnamefont {Pescia}},\
  }\bibfield  {title} {\enquote {\bibinfo {title} {Observation of stripe
  mobility in a dipolar frustrated ferromagnet},}\ }\href@noop {} {\bibfield
  {journal} {\bibinfo  {journal} {Phys. Rev. Lett.}\ }\textbf {\bibinfo
  {volume} {96}},\ \bibinfo {pages} {047212} (\bibinfo {year}
  {2006})}\BibitemShut {NoStop}%
\bibitem [{\citenamefont {Schmalian}, \citenamefont {Wolynes},\ and\
  \citenamefont {Wu}(2003)}]{Schmalian2003}%
  \BibitemOpen
  \bibfield  {author} {\bibinfo {author} {\bibfnamefont {J.}~\bibnamefont
  {Schmalian}}, \bibinfo {author} {\bibfnamefont {P.~G.}\ \bibnamefont
  {Wolynes}}, \ and\ \bibinfo {author} {\bibfnamefont {S.}~\bibnamefont {Wu}},\
  }\bibfield  {title} {\enquote {\bibinfo {title} {Comment on `` the nature of
  slow dynamics in a minimal model of frustration limited domain'' by {PL}
  {G}eissler and {DR} {R}eichman, cond-mat/0304254},}\ }\href@noop {}
  {\bibfield  {journal} {\bibinfo  {journal} {arXiv:cond-mat/0305420}\ }
  (\bibinfo {year} {2003})}\BibitemShut {NoStop}%
\bibitem [{\citenamefont {Haw}(2010)}]{Haw2010}%
  \BibitemOpen
  \bibfield  {author} {\bibinfo {author} {\bibfnamefont {M.~D.}\ \bibnamefont
  {Haw}},\ }\bibfield  {title} {\enquote {\bibinfo {title} {Growth kinetics of
  colloidal chains and labyrinths},}\ }\href {\doibase
  10.1103/PhysRevE.81.031402} {\bibfield  {journal} {\bibinfo  {journal} {Phys.
  Rev. E}\ }\textbf {\bibinfo {volume} {81}},\ \bibinfo {pages} {031402}
  (\bibinfo {year} {2010})}\BibitemShut {NoStop}%
\bibitem [{\citenamefont {Zhang}\ and\ \citenamefont {Wang}(2006)}]{Zhang2006}%
  \BibitemOpen
  \bibfield  {author} {\bibinfo {author} {\bibfnamefont {C.-Z.}\ \bibnamefont
  {Zhang}}\ and\ \bibinfo {author} {\bibfnamefont {Z.-G.}\ \bibnamefont
  {Wang}},\ }\bibfield  {title} {\enquote {\bibinfo {title} {Random isotropic
  structures and possible glass transitions in diblock copolymer melts},}\
  }\href@noop {} {\bibfield  {journal} {\bibinfo  {journal} {Phys. Rev. E}\
  }\textbf {\bibinfo {volume} {73}},\ \bibinfo {pages} {031804} (\bibinfo
  {year} {2006})}\BibitemShut {NoStop}%
\bibitem [{\citenamefont {Tarzia}\ and\ \citenamefont
  {Coniglio}(2007)}]{Tarzia2007}%
  \BibitemOpen
  \bibfield  {author} {\bibinfo {author} {\bibfnamefont {M.}~\bibnamefont
  {Tarzia}}\ and\ \bibinfo {author} {\bibfnamefont {A.}~\bibnamefont
  {Coniglio}},\ }\bibfield  {title} {\enquote {\bibinfo {title} {Lamellar
  order, microphase structures, and glassy phase in a field theoretic model for
  charged colloids},}\ }\href@noop {} {\bibfield  {journal} {\bibinfo
  {journal} {Phys. Rev. E}\ }\textbf {\bibinfo {volume} {75}},\ \bibinfo
  {pages} {011410} (\bibinfo {year} {2007})}\BibitemShut {NoStop}%
\bibitem [{\citenamefont {Li}\ \emph {et~al.}(2014)\citenamefont {Li},
  \citenamefont {Nealey}, \citenamefont {de~Pablo},\ and\ \citenamefont
  {M\"uller}}]{Li2014}%
  \BibitemOpen
  \bibfield  {author} {\bibinfo {author} {\bibfnamefont {W.}~\bibnamefont
  {Li}}, \bibinfo {author} {\bibfnamefont {P.~F.}\ \bibnamefont {Nealey}},
  \bibinfo {author} {\bibfnamefont {J.~J.}\ \bibnamefont {de~Pablo}}, \ and\
  \bibinfo {author} {\bibfnamefont {M.}~\bibnamefont {M\"uller}},\ }\bibfield
  {title} {\enquote {\bibinfo {title} {Defect removal in the course of directed
  self-assembly is facilitated in the vicinity of the order-disorder
  transition},}\ }\href {\doibase 10.1103/PhysRevLett.113.168301} {\bibfield
  {journal} {\bibinfo  {journal} {Phys. Rev. Lett.}\ }\textbf {\bibinfo
  {volume} {113}},\ \bibinfo {pages} {168301} (\bibinfo {year}
  {2014})}\BibitemShut {NoStop}%
\bibitem [{\citenamefont {Tarzia}\ and\ \citenamefont
  {Coniglio}(2006)}]{Tarzia2006}%
  \BibitemOpen
  \bibfield  {author} {\bibinfo {author} {\bibfnamefont {M.}~\bibnamefont
  {Tarzia}}\ and\ \bibinfo {author} {\bibfnamefont {A.}~\bibnamefont
  {Coniglio}},\ }\bibfield  {title} {\enquote {\bibinfo {title} {Pattern
  formation and glassy phase in the ${\ensuremath{\phi}}^{4}$ theory with a
  screened electrostatic repulsion},}\ }\href@noop {} {\bibfield  {journal}
  {\bibinfo  {journal} {Phys. Rev. Lett.}\ }\textbf {\bibinfo {volume} {96}},\
  \bibinfo {pages} {075702} (\bibinfo {year} {2006})}\BibitemShut {NoStop}%
\bibitem [{\citenamefont {Uneyama}(2007)}]{Uneyama2007}%
  \BibitemOpen
  \bibfield  {author} {\bibinfo {author} {\bibfnamefont {T.}~\bibnamefont
  {Uneyama}},\ }\bibfield  {title} {\enquote {\bibinfo {title} {Density
  functional simulation of spontaneous formation of vesicle in block copolymer
  solutions},}\ }\href@noop {} {\bibfield  {journal} {\bibinfo  {journal} {J.
  Chem. Phys.}\ }\textbf {\bibinfo {volume} {126}},\ \bibinfo {eid} {114902}
  (\bibinfo {year} {2007})}\BibitemShut {NoStop}%
\bibitem [{\citenamefont {Groot}\ and\ \citenamefont
  {Warren}(1997)}]{Groot1997}%
  \BibitemOpen
  \bibfield  {author} {\bibinfo {author} {\bibfnamefont {R.~D.}\ \bibnamefont
  {Groot}}\ and\ \bibinfo {author} {\bibfnamefont {P.~B.}\ \bibnamefont
  {Warren}},\ }\bibfield  {title} {\enquote {\bibinfo {title} {Dissipative
  particle dynamics: Bridging the gap between atomistic and mesoscopic
  simulation},}\ }\href@noop {} {\bibfield  {journal} {\bibinfo  {journal} {J.
  Chem. Phys.}\ }\textbf {\bibinfo {volume} {107}},\ \bibinfo {pages} {4423}
  (\bibinfo {year} {1997})}\BibitemShut {NoStop}%
\bibitem [{\citenamefont {Riesch}, \citenamefont {Radons},\ and\ \citenamefont
  {Magerle}(2014)}]{Riesch2014}%
  \BibitemOpen
  \bibfield  {author} {\bibinfo {author} {\bibfnamefont {C.}~\bibnamefont
  {Riesch}}, \bibinfo {author} {\bibfnamefont {G.}~\bibnamefont {Radons}}, \
  and\ \bibinfo {author} {\bibfnamefont {R.}~\bibnamefont {Magerle}},\
  }\bibfield  {title} {\enquote {\bibinfo {title} {Aging of orientation
  fluctuations in stripe phases},}\ }\href {\doibase
  10.1103/PhysRevE.90.052101} {\bibfield  {journal} {\bibinfo  {journal} {Phys.
  Rev. E}\ }\textbf {\bibinfo {volume} {90}},\ \bibinfo {pages} {052101}
  (\bibinfo {year} {2014})}\BibitemShut {NoStop}%
\bibitem [{\citenamefont {Lindquist}, \citenamefont {Jadrich},\ and\
  \citenamefont {Truskett}(2016)}]{Lindquist2016}%
  \BibitemOpen
  \bibfield  {author} {\bibinfo {author} {\bibfnamefont {B.~A.}\ \bibnamefont
  {Lindquist}}, \bibinfo {author} {\bibfnamefont {R.~B.}\ \bibnamefont
  {Jadrich}}, \ and\ \bibinfo {author} {\bibfnamefont {T.~M.}\ \bibnamefont
  {Truskett}},\ }\bibfield  {title} {\enquote {\bibinfo {title} {Assembly of
  nothing: Equilibrium fluids with designed structured porosity},}\ }\href@noop
  {} {\bibfield  {journal} {\bibinfo  {journal} {Soft matter}\ }\textbf
  {\bibinfo {volume} {12}},\ \bibinfo {pages} {2663--2667} (\bibinfo {year}
  {2016})}\BibitemShut {NoStop}%
\bibitem [{\citenamefont {Lindquist}\ \emph {et~al.}(2017)\citenamefont
  {Lindquist}, \citenamefont {Dutta}, \citenamefont {Jadrich}, \citenamefont
  {Milliron},\ and\ \citenamefont {Truskett}}]{Lindquist2017}%
  \BibitemOpen
  \bibfield  {author} {\bibinfo {author} {\bibfnamefont {B.~A.}\ \bibnamefont
  {Lindquist}}, \bibinfo {author} {\bibfnamefont {S.}~\bibnamefont {Dutta}},
  \bibinfo {author} {\bibfnamefont {R.~B.}\ \bibnamefont {Jadrich}}, \bibinfo
  {author} {\bibfnamefont {D.}~\bibnamefont {Milliron}}, \ and\ \bibinfo
  {author} {\bibfnamefont {T.~M.}\ \bibnamefont {Truskett}},\ }\bibfield
  {title} {\enquote {\bibinfo {title} {Interactions and design rules for
  assembly of porous colloidal mesophases},}\ }\href@noop {} {\bibfield
  {journal} {\bibinfo  {journal} {Soft Matter}\ } (\bibinfo {year}
  {2017})}\BibitemShut {NoStop}%
\bibitem [{\citenamefont {Towns}\ \emph {et~al.}(2014)\citenamefont {Towns},
  \citenamefont {Cockerill}, \citenamefont {Dahan}, \citenamefont {Foster},
  \citenamefont {Gaither}, \citenamefont {Grimshaw}, \citenamefont {Hazlewood},
  \citenamefont {Lathrop}, \citenamefont {Lifka}, \citenamefont {Peterson},
  \citenamefont {Roskies}, \citenamefont {Scott},\ and\ \citenamefont
  {Wilkins-Diehr}}]{6866038}%
  \BibitemOpen
  \bibfield  {author} {\bibinfo {author} {\bibfnamefont {J.}~\bibnamefont
  {Towns}}, \bibinfo {author} {\bibfnamefont {T.}~\bibnamefont {Cockerill}},
  \bibinfo {author} {\bibfnamefont {M.}~\bibnamefont {Dahan}}, \bibinfo
  {author} {\bibfnamefont {I.}~\bibnamefont {Foster}}, \bibinfo {author}
  {\bibfnamefont {K.}~\bibnamefont {Gaither}}, \bibinfo {author} {\bibfnamefont
  {A.}~\bibnamefont {Grimshaw}}, \bibinfo {author} {\bibfnamefont
  {V.}~\bibnamefont {Hazlewood}}, \bibinfo {author} {\bibfnamefont
  {S.}~\bibnamefont {Lathrop}}, \bibinfo {author} {\bibfnamefont
  {D.}~\bibnamefont {Lifka}}, \bibinfo {author} {\bibfnamefont {G.~D.}\
  \bibnamefont {Peterson}}, \bibinfo {author} {\bibfnamefont {R.}~\bibnamefont
  {Roskies}}, \bibinfo {author} {\bibfnamefont {J.~R.}\ \bibnamefont {Scott}},
  \ and\ \bibinfo {author} {\bibfnamefont {N.}~\bibnamefont {Wilkins-Diehr}},\
  }\bibfield  {title} {\enquote {\bibinfo {title} {{XSEDE}: Accelerating
  scientific discovery},}\ }\href {\doibase 10.1109/MCSE.2014.80} {\bibfield
  {journal} {\bibinfo  {journal} {Computing in Science Engineering}\ }\textbf
  {\bibinfo {volume} {16}},\ \bibinfo {pages} {62--74} (\bibinfo {year}
  {2014})}\BibitemShut {NoStop}%
\end{thebibliography}%
\end{document}